\documentclass[a4paper,12pt]{article}
\usepackage[a4paper, bindingoffset=0.2in, left=1in, right=1in, top=1in, bottom=1in, footskip=.25in]{geometry}

\usepackage{times}
\usepackage{amsmath,amssymb,amsthm}
\usepackage{graphicx}
\usepackage{subcaption}
\usepackage{authblk}
\usepackage[nottoc]{tocbibind}
\usepackage{setspace}
\usepackage{bbm}
\usepackage{hyperref}
\usepackage{xcolor}
\hypersetup{
  colorlinks   = true, 
  urlcolor     = red, 
  linkcolor    = blue, 
  citecolor   = blue 
}
\usepackage[labelsep=period, labelfont=sc]{caption}
\usepackage{tabularx,ragged2e,booktabs}
\usepackage{sectsty}

\usepackage{caption}
\usepackage{lineno}

\usepackage{enumitem}

\newcommand{\vect}[1]{\boldsymbol{#1}}

\title{Minimum Description Length codes are critical}

\author{Ryan Cubero, Matteo Marsili and Yasser Roudi}
\date{\footnotesize \today}  

\begin{document}
\maketitle

\begin{abstract}
In the Minimum Description Length (MDL) principle, learning from the data is equivalent to an optimal coding problem. We show that the codes that achieve optimal compression in MDL are critical in a very precise sense. First, when they are taken as generative models of samples, they generate samples with broad empirical distributions and with a high value of the relevance, defined as the entropy of the empirical frequencies. These results are derived for different statistical models (Dirichlet model, independent and pairwise dependent spin models, and restricted Boltzmann machines). Second, MDL codes sit precisely at a second order phase transition point where the symmetry between the sampled outcomes is spontaneously broken. The order parameter controlling the phase transition is the coding cost of the samples. The phase transition is a manifestation of the optimality of MDL codes, and it arises because codes that achieve a higher compression do not exist. These results suggest a clear interpretation of the widespread occurrence of statistical criticality as a characterization of samples which are maximally informative on the underlying generative process.
\end{abstract}

\section{Introduction}

\begin{figure}[ht]
\centering
\includegraphics[width=0.8\textwidth]{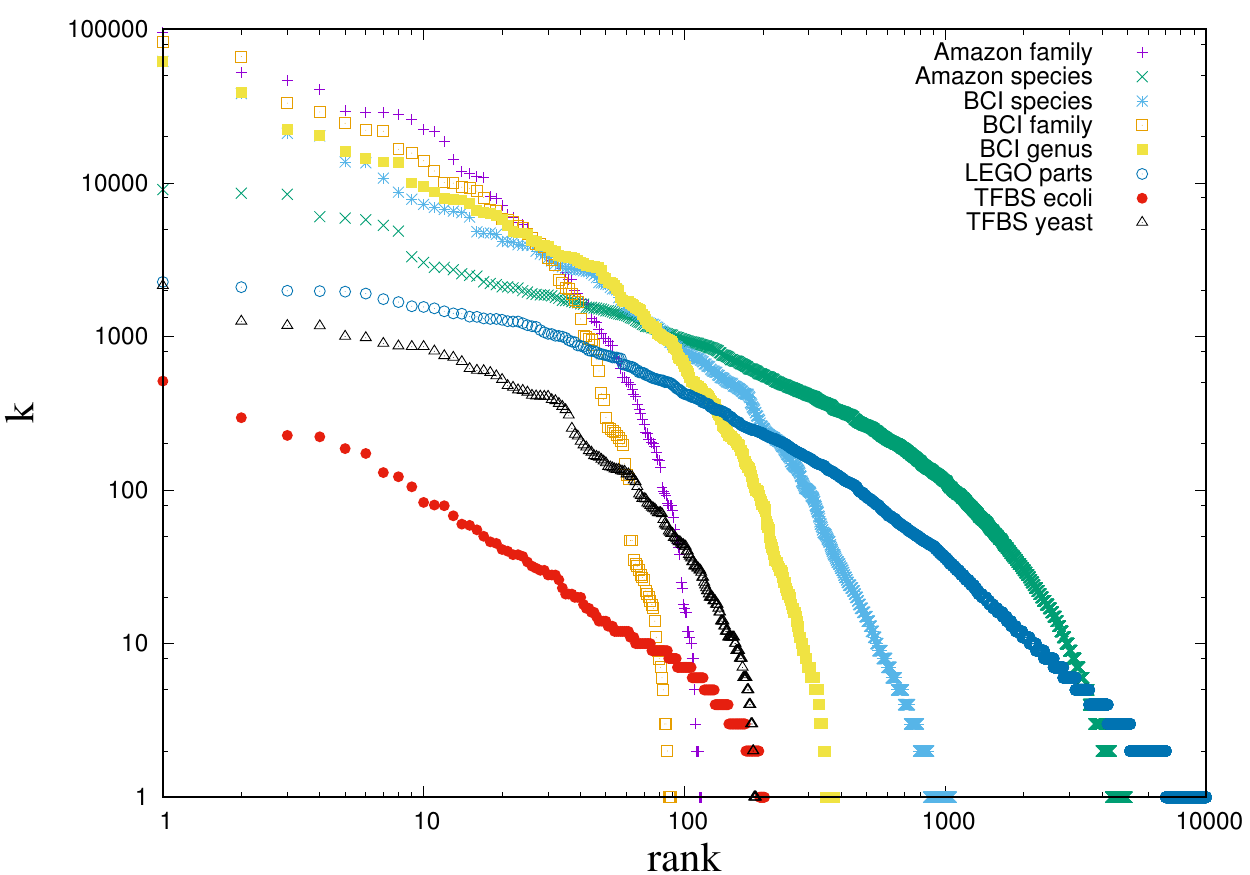}
\caption{\label{fig:empirical} { Rank plot of the frequencies across a broad range of datasets.}
Log-log plots of rank versus frequency from diverse datasets: survey of 4962 species of trees across 116 families sampled from the Amazonian lowlands~\cite{ter2013hyperdominance}, survey of 1053 species of trees across 376 genera and 89 families sampled across a 50 hectare plot in the Barro Colorado Island (BCI), Panama~\cite{condit2012barro}, counts indicating the inclusion of each 13,001 LEGO parts on 2613 distributed toy sets~\cite{rebrickable, mazzolini2018statistics} and the number of genes that are regulated by each of the 203 transcription factors (TFs) in \emph{E. coli}~\cite{gama2015regulondb} and 188 TFs in \emph{S. cerevisiae} (yeast)~\cite{balakrishnan2012yeastmine} through binding with transcription factor binding sites (TFBS).}
\end{figure}

It is not infrequent to find empirical data which exhibits broad frequency distributions in the most disparate domains.
Broad distributions manifest in the fact that if outcomes are ranked in order of decreasing frequency of their occurrence, then the rank frequency plot spans several orders of magnitude on both axes.
Figure \ref{fig:empirical} reports few cases (see caption for details), but many more have been reported in the literature (see e.g.,  \cite{munoz2018colloquium,newman2005power}).
A straight line in the rank plot corresponds to a power law frequency distribution, where the number of outcomes that are observed $k$ times behave as $m_k\sim k^{-\mu-1}$ (with $1/\mu$ being the slope of the rank plot).
Yet, as Figure  \ref{fig:empirical} shows, empirical distributions are not always power laws, even though they are broad nonetheless.
Countless mechanisms have been advanced to explain this behaviour \cite{munoz2018colloquium, newman2005power, bak1996nature, mora2011biological, simini2012universal, schwab2014zipf}.
It has recently been suggested that broad distributions arise from efficient representations, i.e., when the data samples relevant variables, which are those carrying the maximal amount of information on the generative process \cite{marsili2013sampling, haimovici2015criticality, cubero2018minimally}.
Such Maximally Informative Samples (MIS) are those for which the entropy of the frequency with which outcomes occur---called {\em relevance} in
~\cite{haimovici2015criticality, cubero2018minimally}---is maximal at a given resolution, which is measured by the number of bits needed to encode the sample (see Section \ref{sec:relevance}).
MIS exhibit power law distributions with the exponent $\mu$ governing the tradeoff between resolution and relevance \cite{cubero2018minimally}.
This argument for the emergence of broad distributions is independent of any mechanism or model.
A direct way to confirm this claim is to check that samples generated from models that are known to encode efficient representations are actually maximally informative.
In this line, \cite{song2017resolution} found strong evidence that MIS occur in the representations that deep learning extracts from data.
This paper explores the same issue in efficient coding as defined in Minimum Description Length \cite{grunwald2007minimum}. 

Regarding empirical data as a message sent from nature, we expect it to be expressed in an efficient manner if relevant variables are chosen.
This requirement can be made quantitative and precise, in information theoretic terms, following Minimum Description Length (MDL) theory \cite{grunwald2007minimum}.
MDL seeks the optimal encoding of data generated by a parametric model with unknown parameters (see Section~\ref{sec:MDL}).
MDL derives a probability distribution over samples that embodies the requirement of optimal encoding.
This distribution is the Normalized Maximum Likelihood (NML). 
This paper studies the NML as a generative process of samples and studies both its typical and atypical properties.
In a series of cases, we find that samples generated by NMLs are typically close to being maximally informative, in the sense of \cite{cubero2018minimally}, and that their frequency distribution is typically broad.
In addition, we find that NMLs are critical in a very precise sense, because they sit at a second order phase transition that separates typical from atypical behavior. 
More precisely, we find that large deviations, for which the resolution attains atypically low values, exhibit a condensation phenomenon whereby all $N$ points in the sample coincide.
This is consistent with the fact that NML correspond to efficient coding of random samples generated from a model, so that codes achieving higher compression do not exist.
Large deviations enforcing higher compression force parameters to corners of the allowed space where the model becomes deterministic.

The rest of the paper is organized as follows: the rest of the introduction lays the background of what follows by recalling the characterization of samples in terms of resolution and relevance, as in \cite{cubero2018minimally}, and the derivation of NML in MDL, following \cite{grunwald2007minimum}. Section \ref{sec:typical} discusses typical properties of NML and Section \ref{sec:atypical} discusses large deviations of the coding cost. We conclude with a series of remarks on the significance of these results.

\noindent Setting the Scene 

Let $\hat{s} = (s^{(1)}, \ldots, s^{(N)})$ be a sample of $N$ observations, $s^{(i)} \in \chi$, of a system where $\chi$ is a countable finite state space.
We define $k_s$ as the number of observations in $\hat{s}$ for which $s^{(i)}=s$, i.e., the frequency of $s$.
The number of states $s$ that occur $k$ times will be denoted as $m_k$. Both $k_s$ and $m_k$ depend on the sample $\hat{s}$. 

We assume that $\hat{s}$ is generated in a series of independent experiments or observations, all in the same conditions.
This is equivalent to taking $\hat{s}$ as a sequence of $N$ independent draws from an unknown distribution $p(s)$ (i.e., the generative process).

\subsection{Resolution, Relevance and Maximally Informative Samples}\label{sec:relevance} 
The information content of the sample is measured by the number of bits needed to encode a single data point.
This is given by Shannon entropy \cite{cover2012elements}.
Taking the frequency $k_s/N$ as the probability of point $s$, this leads to:

\begin{equation}
\hat{H}[s]=-\sum_{s\in\chi} \frac{k_s}{N}\log \frac{k_s}{N}=-\sum_k \frac{km_k}{N}\log \frac{k}{N},
\label{Hs}
\end{equation}

\noindent where the $\hat{~}$ indicates that the entropy is computed from the empirical frequency.
This quantity specifies the level of detail of the description provided by the variable $s$.
At one extreme, all the data points are equal, i.e., $s^{(i)} = s$, $\forall i=1, \ldots, N$ such that $m_k=0$ for $k=1,\ldots,N-1$ and $m_{k=N}=1$.
With this, 
one finds that $\hat{H}[s]=0$.
On the other extreme, all the data points are different, i.e., $s^{(i)} \neq s^{(j)}$, $\forall~ i \neq j$, such that $m_{k=1}=N$ and $m_{k'} = 0$, $\forall k' > 1$. Hence, 
one finds that $\hat{H}[s]=\log N$.
This is why we call $\hat{H}[s]$ as the {\em resolution}, following \cite{cubero2018minimally}.
The resolution clearly depends on the cardinality of $\chi$.
Only a part of $\hat{H}[s]$ provides information on the generative process $p(s)$ and this is given by the {\em relevance}

\begin{equation}
\label{Hk}
\hat{H}[k] = - \sum_k \frac{km_k}{N} \log \frac{km_k}{N}.
\end{equation}

\noindent A simple argument, which is elaborated in detail in \cite{cubero2018minimally}, is that the empirical frequency $k_s/N$ is the best estimate of $p(s)$, so conditional on $k_s$, the sample does not contain any further information on $p(s)$. 
Note that $k_s$ is a function of $s$, which implies $\hat{H}[s,k] = \hat{H}[s] \ge \hat{H}[k]$.
Therefore, the difference $\hat{H}[s] - \hat{H}[k] = \hat{H}[s|k]$ quantifies the amount of noise the sample contains. 

We call $\hat{s}$ a Maximally Informative Sample (MIS) if $m_k$ is such that the relevance is maximal at a given resolution $\hat{H}[s]=H_0$.
This implies the maximization of the functional

\begin{equation}
\mathcal{F} = \hat{H}[k] + \mu (\hat{H}[s] - H_0) + \lambda \left( \sum_k k m_k - N \right)
\end{equation}

\noindent over $m_k$, where the Lagrange multipliers $\mu$ and $\lambda$ are adjusted to enforce the conditions $\hat{H}[s] = H_0$ and $\sum_k k m_k = N$, respectively.
As shown in \cite{marsili2013sampling,haimovici2015criticality}, MIS exhibit a power law frequency distribution

\begin{equation}
m_k \approx c k^{-1-\mu}
\label{power_law}
\end{equation}

\noindent where $c$ is a normalization constant such that $\sum_k k m_k = N$.
As $H_0$ varies from $0$ to $\log N$, MISs trace a curve in the resolution-relevance plane (see solid lines in Figure \ref{fig:uc_dirichlet},\ref{fig:uc_paramagnet} B, C) with $\mu$ as the negative slope.
As discussed in \cite{song2017resolution,cubero2018minimally}, $\mu$ quantifies the trade-off between resolution and relevance: a decrease in resolution of one bit leads to an increase of $\mu $ bits in relevance.
The point $\mu = 1$, which corresponds to Zipf's law, sets the limit beyond which further reduction in $\hat{H}[s]$ results in lossy compression, because, for $\mu<1$, the increase in $\hat{H}[k]$ cannot compensate the loss in resolution.

\subsection{Minimum Description Length and the Normalized Maximum Likelihood}\label{sec:MDL}
The main insight of MDL is that learning from data is equivalent to data compression \cite{grunwald2007minimum}.
In turn, data compression is equivalent to assigning a probability distribution over the space of samples.
This section provides a brief derivation of this distribution whereas the rest of the paper discuss its typical and atypical properties.
We refer the interested reader to \cite{grunwald2007minimum,grunwald2004tutorial} for a more detailed discussion of MDL. 

From an information theoretic perspective, one can think of the sample, $\hat{s}$, as a message generated by some source (e.g., nature) that we wish to compress as much as possible.
This entails translating $\hat{s}$ in a sequence of bits.
A code is a rule that achieves this for any $\hat{s} \in \chi^N$ and its efficiency depends on whether frequent patterns are assigned short codewords or not.
Conversely, any code implies a distribution $P(\hat{s})$ over the space of samples and the cost of encoding the sample $\hat{s}$ under the code $P$ is given by~\cite{cover2012elements} 

\begin{equation}
E = - \log P(\hat{s})
\label{energy}
\end{equation}

\noindent bits (assuming logarithm base two). Optimal compression is achieved when the code $P$ coincides with the data generating process~\cite{cover2012elements}. 

Consider the situation where the data is generated as independent draws from a parametric model $f(s|\theta)$.
If the value of $\theta$ were known, then the optimal code would be given by $P(\hat{s}) = \prod_i f(s^{(i)} | \theta) \equiv f(\hat{s} | \theta)$.
MDL seeks to derive $P$ in the case where $\theta$ is not known (Indeed, MDL aims at deriving efficient coding under $f$ irrespective of whether $f(s | \theta)$ is the ``true'' generative model or not.
This allows one to compare different models and choose the one providing the most concise description of the data).
This applies, for example, to the situation where $\hat{s}$ is a series of experiments or observation aimed at measuring the parameters $\theta$ of a theory. 

In hind sight, i.e., upon seeing the sample, the best code is $f(\hat{s} | \hat{\theta})$, where $\hat{\theta}(\hat{s})$ is the maximum likelihood estimator for $\theta$, and it depends on the sample $\hat{s}$.
Therefore, one can define the \emph{regret} $\mathcal{R}$, as the additional encoding cost that one needs to spend to encode the sample $\hat{s}$, if one uses the code $P(\hat{s})$ to compress $\hat{s}$, i.e.,

\begin{equation}
\mathcal{R} = - \log P(\hat{s}) - \min_\theta \left[ - \log f(\hat{s} | \theta) \right].
\label{regret}
\end{equation}

\noindent Notice that $\min_\theta \left[ - \log f(\hat{s} | \theta) \right] = - \log f( \hat{s} | \hat{\theta}(\hat{s}))$.
$\mathcal{R}$ is called regret of $P$ relative to $f$ for sample $\hat{s}$ because it depends both on $P$ and on $\hat{s}$. 

MDL derives the optimal code, $\bar{P}(\hat{s})$, that minimizes the regret, assuming that for any $P$ the source produces the worst possible sample~\cite{grunwald2007minimum}.
The solution~\cite{shtarkov1987universal}

\begin{equation}
\bar{P}(\hat{s}) 	= \frac{ f(\hat{s} | \hat{\theta}(\hat{s})) }{ \sum_{\hat{x} \in \chi^N} f(\hat{x} | \hat{\theta}(\hat{x})) }.
\label{NML_prob}
\end{equation}

\noindent is called the Normalized Maximum Likelihood (NML).
The optimal regret is given by

\begin{equation}
\bar{\mathcal{R}} = \log \sum_{\hat{s} \in \chi^N} f(\hat{s} | \hat{\theta}(\hat{s}))
\label{UC_regret}
\end{equation}

\noindent which is known in MDL as the \emph{parametric complexity} ({Notice that $e^{\bar{\mathcal{R}}}$ can be seen as a partition sum. Hence, throughout the paper, we shall refer to the parametric complexity as the \emph{UC partition function}.}).
For models in the exponential family, Rissanen showed that the parametric complexity is asymptotically given by~\cite{rissanen1996fisher}

\begin{equation}
\bar{\mathcal{R}} \simeq \frac{k}{2} \log \frac{N}{2 \pi} + \log \int \sqrt{ \det I(\theta) } d\theta + \mathcal{O}(1)
\label{parametric_complexity}
\end{equation} 

\noindent where $I(\theta)$ is the Fisher information matrix with the matrix elements defined by an expectation $I_{ij}(\theta) = - \langle \frac{\partial^2 \log f( s | \theta)}{\partial \theta_i \partial \theta_j} \rangle_\theta$ over the parametric model $f( {s} | \theta)$ (see Appendix \ref{saddle} for a simple derivation).
The NML code is a {\em universal code} because it achieves a compression per data point which is as good as the compression that would be achieved with the optimal choice of $\theta$ when one has large enough samples.
This is easy to see, because the regret $\bar{\mathcal{R}}/N$ per data point vanishes in the limit $N\to\infty$, hence the NML code achieves the same compression as $f(\hat{s}|\hat \theta)$. 

Notice also that the optimal regret, $\bar{\mathcal{R}}$, in Equation \eqref{UC_regret} is independent of the sample $\hat{s}$.
It indeed provides a measure of complexity of the model $f$ that can be used in model selection schemes.
For exponential families, MDL procedure
penalizes
models with a cost which equals the one obtained in Bayesian model selection~\cite{balasubramanian2005mdl} under a Jeffreys prior.
Indeed, considering $\bar P(\hat{s})$ as a generative model for samples, one can show that the induced distribution on $\theta$ is given by Jeffreys prior (see Appendix \ref{saddle}).

\section{Results}

\subsection{NML Codes Provide Efficient Representations}
\label{sec:typical}

In this section we consider $\bar P$ as a generative model for samples and we investigate its typical properties for some representative statistical models.

\subsubsection{Dirichlet Model}

Let us start by considering the Dirichlet model distribution $f(s|\theta)=\theta_s$, $\forall s \in \chi$. The parameters $\theta_s\ge 0$ are constrained by the normalization condition $\sum_{s\in \chi}\theta_s=1$. 
Let $S=|\chi|$ denote the cardinality of $\chi$ and define, for convenience, $\rho=N/S$ as the average number of points per state.
Because each observation is mutually independent, the likelihood of a sample $\hat{s}$ given ${\theta}=(\theta_1, \ldots, \theta_S)$ can be written as

\begin{equation}
f( \hat{s} | \theta ) = \prod_{s \in \chi} \theta_s^{k_s},
\end{equation}

\noindent where $k_s$ is the number of times that the state $s$ occurs in the sample $\hat{s}$.
From here, it can be seen that $\hat\theta_s = k_s/N$ is the maximum likelihood estimator for $\theta_s$.
Thus, the universal code for the Dirichlet model can now be constructed as

\begin{equation}
\bar{P}(\hat{s}) = e^{-\bar{\mathcal{R}}} \prod_{s \in \chi} \left( \frac{k_s}{N} \right)^{k_s}
\label{pofs_dirichlet}
\end{equation}

\noindent which can be read as saying that for each $s$, the code needs $-k_s \log(k_s/N) +\bar{\mathcal{R}}/N$ bits.
In terms of the frequencies, $\{ k_1, \ldots, k_S \}$, the universal codes can be written as

\begin{equation}
\bar{P}( k_1, \ldots, k_S ) = e^{-\bar{\mathcal{R}}} \frac{N!}{\prod_{s \in \chi} k_s!} \prod_{s \in \chi} \left( \frac{k_s}{N} \right)^{k_s} \delta \left( \sum_{s \in \chi} k_s - N \right)
\label{pofk_dirichlet}
\end{equation}

\noindent wherein the multinomial coefficient, $\frac{N!}{\prod_{s \in \chi} k_s!}$, counts the number of samples with a given frequency profile $k_1, \ldots, k_S$.
In order to compute the optimal regret $\bar{\mathcal{R}}$, we have to evaluate the partition function

\begin{align}
e^{\bar{\mathcal{R}}} &= \frac{N!}{N^N e^{-N}} \int_{-\pi}^{\pi} \frac{d \mu}{2 \pi} e^{i \mu N} \left[\sum_{k_1=0}^\infty \frac{k_1^{k_1}e^{- k_1} e^{- i\mu k_1}}{k_1!} \right] \cdots \left[\sum_{k_S=0}^\infty \frac{k_S^{k_S}e^{- k_S} e^{- i\mu k_S}}{k_S!} \right] \\
&= \frac{N!}{N^N e^{-N}} \int_{-\pi}^{\pi} \frac{d \mu}{2 \pi} e^{i \mu N} \left[ \mathcal{N}(i\mu) \right]^S \\
&\simeq \sqrt{2\pi N} \int_{-\pi}^{\pi} \frac{d \mu}{2 \pi} e^{S \Phi(i \mu)} \label{dirichlet_integral_saddlepoint}
\end{align}

\noindent where

\begin{equation}
\Phi(z) = \rho z + \log \mathcal{N}(z)
\label{phiofz}
\end{equation}

\noindent and

\begin{equation}
\mathcal{N}(z) = \sum_{k=0}^\infty \frac{k^{k}e^{- (1+z)k} }{k!}.
\label{qofk_norm}
\end{equation}

The integral in Equation \eqref{dirichlet_integral_saddlepoint} is dominated by the value where the function $\phi$ attains its saddle point value $z^*(\rho)$, which is given by the condition 

\begin{equation}
\frac{d\Phi}{dz}  = \rho-\langle k\rangle_z =0
\end{equation}

\noindent where the average $\langle\ldots\rangle_z$ is taken with respect to the distribution

\begin{equation}
q( k | z ) = \frac{1}{\mathcal{N}(z)} \frac{k^{k}e^{- (1+z)k} }{k!}.
\label{qofk_dirichlet}
\end{equation}

\noindent Gaussian integration around the saddle point leads then to

\begin{equation}
e^{\bar{\mathcal{R}}} \simeq
\sqrt{\rho}
\frac{e^{S \Phi(z^*(\rho))}}{\sqrt{ \langle k^2 \rangle_{z^*} - \langle k \rangle_{z^*}^2 }}
\label{dirichlet_regretA}
\end{equation}

\noindent where we used the identity $\Phi''(z) = - \left[ \langle k^2 \rangle_{z} - \langle k \rangle_{z}^2 \right]$.

The distribution Equation (\ref{pofk_dirichlet}) can also be written introducing the Fourier representation of the delta function 

\begin{equation}
\bar{P}( k_1, \ldots, k_S ) = \frac{N! e^{-\bar{\mathcal{R}}}}{N^N e^{-N}} \int_{-\pi}^{\pi} \frac{d \mu}{2 \pi} e^{i \mu N} \prod_{s \in \chi} \frac{k_s^{k_s} e^{-(1+i\mu)k_s}}{k_s!}.
\label{pofk_factorization}
\end{equation}

\noindent For typical sequences $k_1,\ldots,k_S$, the integral is also dominated by the value $\mu=-iz^*(\rho)$ that dominates  Equation (\ref{dirichlet_integral_saddlepoint}), which means that the distribution factorizes as

\begin{equation}
\label{eq:pq}
\bar{P}( k_1, \ldots, k_S )\simeq \prod_{s\in\chi}q(k_s|z^*).
\end{equation}

\begin{figure}[!ht]
\centering
\includegraphics[width=0.95\textwidth]{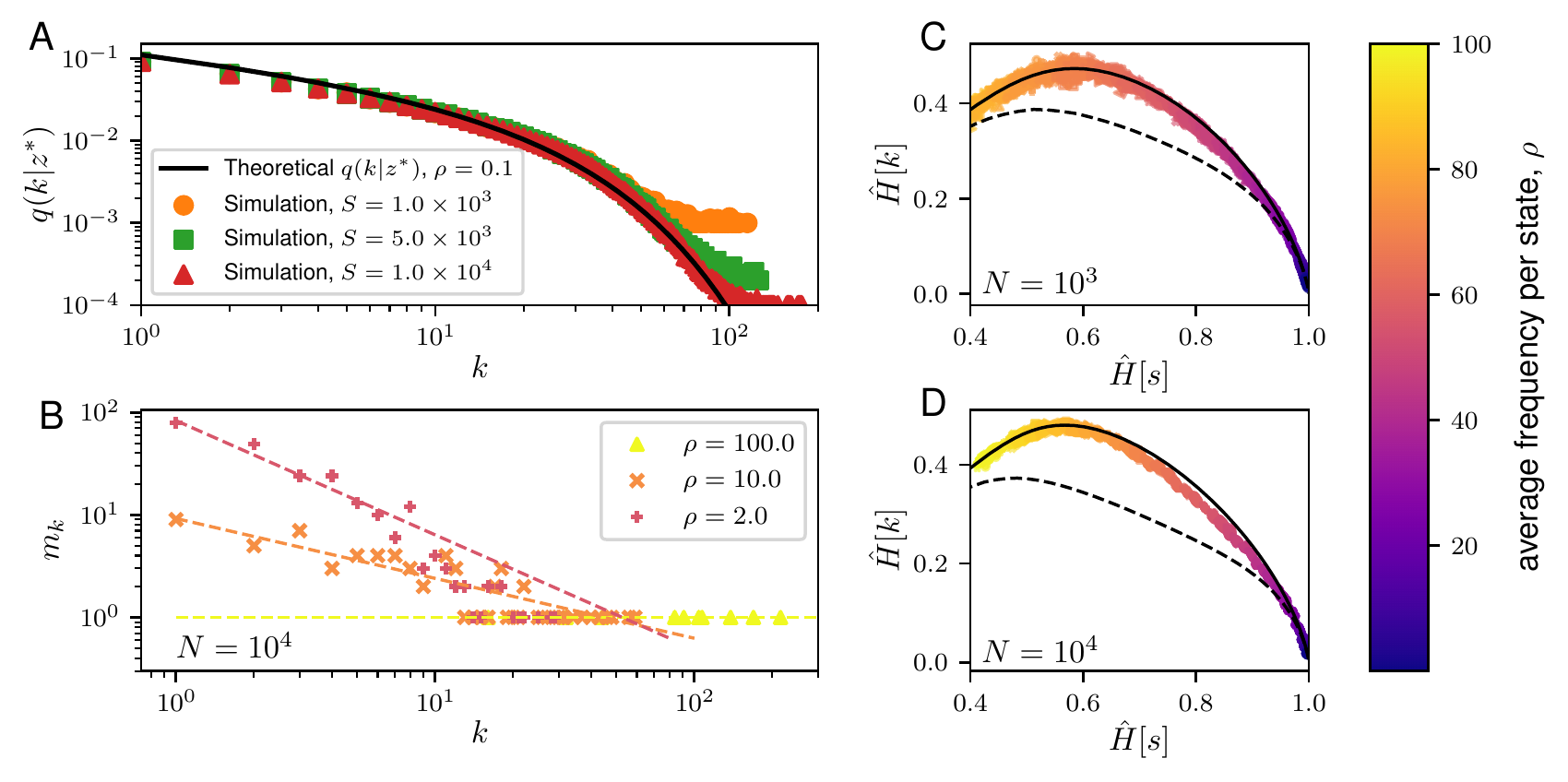}
\caption{\label{fig:uc_dirichlet} {Properties of the typical samples generated from the NML of the Dirichlet model.}
({\bf A}) A plot showing the frequency distribution of the typical samples of the Dirichlet NML code.
Given $S$, the cardinality of the state space, $\chi$, with $S=1.0 \times 10^3$ (orange dots), $5.0 \times 10^3$ (green squares), and $1.0 \times 10^4$ (red triangles), we compute the average frequency distribution across $100$ generated samples from the Dirichlet NML of size $N=10S$ such that the average frequency per state, $\rho$, is fixed.
This is compared against the theoretical calculations (solid black line) for $q(k | z^*)$ in Equation \eqref{qofk_dirichlet}.
({\bf B}) Plot showing the degeneracy, $m_k$, of the frequencies, $k$, in a representative typical sample of length $N=10^3$ generated from the Dirichlet NML code with average frequencies per spike: $\rho=100$ (yellow triangle), $\rho=10$ (orange x-mark) and $\rho=2$ (red cross).
The corresponding dashed lines depict the best-fit line.
({\bf C}--{\bf D}) Plots of  $\hat{H}[s]$ versus $\hat{H}[k]$ for the typical samples of the Dirichlet NML code.
For a fixed size of the data, $N$ ($N=10^3$ in C and $N=10^4$ in D), we have drawn $100$ samples from the Dirichlet NML code varying $\rho$, ranging from $2$ to $100$.
The results are compared against the $\hat{H}[k]$ and $\hat{H}[s]$ for maximally informative samples (MIS, solid black line) and random samples (dashed black lines).
For the MIS, the theoretical lower bound is reported~\cite{haimovici2015criticality}.
For the random samples, we compute the averages of $\hat{H}[s]$ and $\hat{H}[k]$ over $10^7$ realizations of random distributions of $N$ balls in $L$ boxes, with $L$ ranging from $2$ to $10^7$.
Here, each box corresponds to one state $s = 1,\ldots,L$ and $k_s$ is the number of balls in box $s$.
Note that all the calculated values for $\hat{H}[k]$ and $\hat{H}[s]$ are normalized by $\log N$.}
\end{figure}

\noindent This means that the NML is, to a good approximation, equivalent to $S$ independent draws from the distribution $q(k|z^*)$ or, equivalently, that the distribution $q( k | z^* )$ is the one that characterizes typical samples.
This is fully confirmed by Figure  \ref{fig:uc_dirichlet}A, which compares $q(k|z^*)$ with the empirical distribution of $k_s$ drawn from $\bar P$.
For large $k$, we find $q(k|z^*)\sim e^{-z^*k}/\sqrt{k}$, which shows that the distribution of frequencies is broad, with a cutoff at $1/z^*$. 
This underlying broad distribution is confirmed by Figure~ \ref{fig:uc_dirichlet}B which shows the dependence of the degeneracy $m_k$ with the frequency $k$.

In the regime where $\rho \gg 1$ and $k$ is large, the cutoff extends to large values of $k$ and we find $z^*(\rho) \simeq \frac{1}{2 \rho}$ (see Appendix \ref{dirichlet_partition}).
In addition, the parametric complexity can be computed explicitly via Equation (\ref{parametric_complexity}) in this regime, with the result 

\begin{equation}
\bar{\mathcal{R}}\simeq
\frac{S}{2} (1 + \log \rho) - \frac{1}{2} \log (2 \rho)
, \qquad \rho\gg 1.
\end{equation}

The coding cost of a typical sample is given by

\begin{align}
E &= - \log \bar{P}(\hat{s}) \\
   &= - \sum_{s \in \chi} k_s \log \frac{k_s}{N} + \bar{\mathcal{R}} \\
   &= N\hat H[s] +\bar{\mathcal{R}}.\label{codingcostDirichlet}
\end{align}

\noindent The number of samples with encoding cost $E$ can be computed in the following way.
The number of samples that correspond to a given degeneracy $m_k$ of the states that occurs $k_s=k$ times in $\hat{s}$, is given by 

\begin{equation}
\frac{N!}{\prod_k (k m_k)!}.
\end{equation}

\noindent Therefore, the number of samples with coding cost $E$ is

\begin{align}
W(E) &= \sum_{\{m_k\}\in\mathcal{M}(E)} \frac{N!}{\prod_k (k m_k)!}\\
         &= \sum_{\{m_k\}\in\mathcal{M}(E)} e^{\log N! - \sum_k \log (k m_k)!} \\
         &\sim \sum_{\{m_k\}\in\mathcal{M}(E)} e^{N\hat H[k]},\qquad \rho\gg 1 \label{}
\end{align}

\noindent where $\mathcal{M}(E)$ is the set of all sequences $\{m_k\}$ that are consistent with samples in $\chi^N$ and satisfy Equation~ (\ref{codingcostDirichlet}). 
The last expression assumes $\log M!\simeq M\log M -M$, which is reasonable for $M=km_k\gg 1$, i.e., when $\rho\gg 1$.
In this regime we expect the sum over $\mathcal{M}(E)$ to be dominated by samples with maximal $\hat H[k]$.
Indeed, Figure \ref{fig:uc_dirichlet}C,D   show that samples drawn from $\bar P$ achieve values of $\hat H[k]$ close to the theoretical maximum, especially in the region $\rho\gg 1$.

\subsubsection{A Model of Independent Spins}

In order to corroborate our results for the Dirichlet model, we study the properties of the universal codes for a model of independent spins, i.e., a paramagnet.
For a single spin, $s = \pm 1$, in a  local field $h$, the probability distribution is given by 

\begin{equation}
P( s | h ) = \frac{e^{ s h }}{2 \cosh h}.
\end{equation}

\noindent Thus for a sample $\hat{s}$ of size $N$,

\begin{equation}
P( \hat{s} \vert h ) = e^{\left[ Nmh - N \log ( 2 \cosh h ) \right]}
\end{equation}

\noindent where $m = \frac{1}{N} \sum_{i=1}^M s^{(i)}$ is the local magnetization.
The maximum likelihood estimate for $h$ is $\hat h(m)=\tanh^{-1}m$, hence the universal code for a single spin can be written as

\begin{equation}
\bar{P} ( \hat{s} )= e^{ N \left[ m h(m) - \log \left( 2 \cosh h(m) \right) \right] -\bar{\mathcal{R}}}
\end{equation}

\noindent where $\bar{\mathcal{R}}\simeq \frac{1}{2}\log\frac{\pi N}{2}$ (see Appendix \ref{paramagnet_partition}).
Note that a sample with a magnetization $m$ can be realized by considering the permutation of the up-spins ($s=1$, where there are $\ell = \frac{N+Nm}{2}$ of such spins) and the permutation of the down-spins ($s=-1$, where there are $N-\ell$ of such spins).
Consequently, the magnetization for samples drawn from $\bar P$ has a broad distribution given by the arcsin law
(see Appendix \ref{paramagnet_partition})

\begin{align}
\bar{P}(m) &= \binom{N}{\frac{N-Nm}{2}} e^{ N \left[ m \tanh^{-1} m - \log \left( 2 \cosh \left( \tanh^{-1} m \right) \right)\right] -\bar{\mathcal{R}}}\label{UC_paramagnet_pofm}\\
&\simeq \frac{1}{\pi\sqrt{1-m^2}},\qquad m\in [-1,1].
\end{align}

It is straightforward to see that the model of a single spin is equivalent to a Dirichlet model with two states $\chi=\{-1,+1\}$.
In terms of the number $\ell$ of up-spins, using $m = \frac{2\ell - N}{N}$, the NML for a single spin can be written as

\begin{align}
\bar{P}(\ell) = e^{-\bar{\mathcal{R}}}
\binom{N}{\ell} \left( \frac{\ell}{N} \right)^{\ell} \left( 1 - \frac{\ell}{N} \right)^{N-\ell}.
\end{align}

\noindent The NML for a paramagnet with $n$ independent spins reads as
\begin{equation}
\bar{P}(\ell_1, \ldots, \ell_n) = e^{-n\bar{\mathcal{R}}}
\prod_{i=1}^n \binom{N}{\ell_i} \left( \frac{\ell_i}{N} \right)^{\ell_i} \left( 1 - \frac{\ell_i}{N} \right)^{N-\ell_i}.
\label{UC_paramagnet}
\end{equation}

\begin{figure}[ht]
\centering
\includegraphics[width=1.0\textwidth]{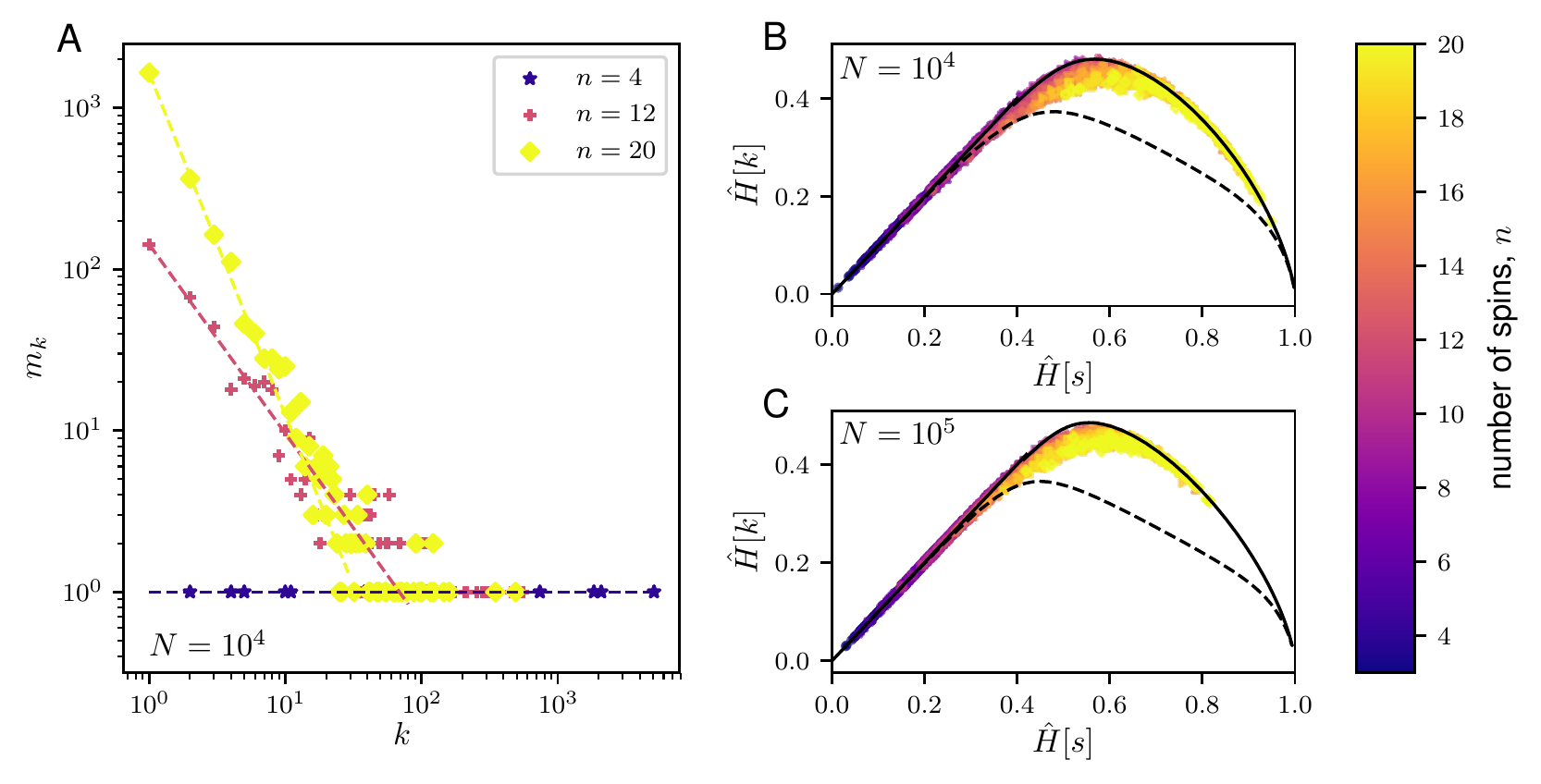}
\caption{\label{fig:uc_paramagnet} { Properties of typical samples for the NML codes of the paramagnet}. 
({\bf A}) Plots showing the degeneracy, $m_k$, of the frequencies, $k$, in a representative typical sample of length $N=10^4$ generated from the NML of a paramagnet with different number of independent spins: $n=4$ (blue star), $n=12$ (red cross) and $n=20$ (yellow diamond).
The corresponding dashed lines depict the best-fit line.
({\bf B}--{\bf C}) Plots of the $\hat{H}[k]$ versus $\hat{H}[s]$ of the typical samples generated from the paramagnet NML code for varying sizes of the data, $N=10^4$ ({\bf B}) and $N=10^5$ ({\bf C}), and for varying number of spins, $n$, ranging from 3 to 20.
Given $N$ and $n$, we compute the $\hat{H}[k]$ and $\hat{H}[s]$ over $100$ realizations of the NML code of a paramagnet.
The results are compared against the $\hat{H}[k]$ and $\hat{H}[s]$ for maximally informative samples (solid black line) and random samples (dashed black line) as described in Figure \ref{fig:uc_dirichlet}.
Note that all the calculated $H[k]$ and $H[s]$ are normalized by $\log N$.}
\end{figure}

Figure \ref{fig:uc_paramagnet} reports the properties of the typical samples of the NML of a paramagnet.
We observed that the frequency distribution of typical samples is broad (Figure \ref{fig:uc_paramagnet}A) and that 
typical samples attain values of $H[k]$ very close to the maximum for a given value of $\hat H[s]$ (Figure \ref{fig:uc_paramagnet}B,C).
As the size $N$ of data  increases, the NML enters the well-sampled regime where $\hat H[k] \simeq \hat H[s]$, indicating that the data processing inequality~\cite{cover2012elements} is saturated.
In this regime, typical samples are those which maximize the entropy $\hat H[s]$.

\subsubsection{Sherrington-Kirkpatrick Model}

In the following sections, we extend our findings to systems of interacting variables (graphical models) and discuss the properties of typical samples drawn from the corresponding NML distribution.
We shall first consider models in which the observed variables are interacting either directly (Sherrington-Kirkpatrick model) and then restricted Boltzmann machines, where the variables interact indirectly through hidden variables.

In this section, $s=(s_1,\ldots,s_n)$ is a configuration of $n$ spins $s_i\in\{\pm 1\}$. 
In the Sherrington-Kirkpatrick (SK) model, the distribution of $s$, considers all interactions up to two-body 

\begin{align}
P( {s} | \vect{J}, \vect{h} ) = \frac{1}{Z(\vect{J}, \vect{h})} \exp \left[ \sum_i h_i s_i + \sum_{i<j} J_{ij} s_i s_j \right], \qquad {s}=(s_1,\ldots, s_n)
\end{align}

\noindent where the partition function

\begin{equation}
Z(\vect{J}, \vect{h})= \sum_{s_1 = \pm 1} \cdots \sum_{s_n = \pm 1} \exp \left[ \sum_i h_i s_i + \sum_{i<j} J_{ij} s_i s_j \right]
\end{equation}

\noindent is a normalization constant which depends on the pairwise couplings, $\vect{J}$ with $J_{ij}=J_{ji}$ being the coupling strength between $s_i$ and $s_j$, and external local fields, $\vect{h}$.
Thus, given a sample, $\hat{{s}} = ( {s}^{(1)}, \ldots, {s}^{(N)} )$ of $N$ observations, the likelihood reads as

\begin{align}
P( \hat{{s}} | \vect{J}, \vect{h} ) = \exp \left[ N \sum_i h_i m_i + N \sum_{i<j} J_{ij} c_{ij} - N \log Z(\vect{J}, \vect{h}) \right].
\end{align}

\noindent where $m_i = \frac{1}{N} \sum_{l=1}^N s_i^{(l)}$ and $c_{ij} = \frac{1}{N} \sum_{l=1}^N s_i^{(l)}s_j^{(l)}$ are the magnetization and pairwise correlation respectively. 
Note that all the needed information about the SK model is encapsulated in the free energy, $\phi(\vect{J},\vect{h}) = \log Z(\vect{J}, \vect{h})$.
Indeed, the maximum likelihood estimators for the couplings, $\hat{\vect{J}}$, and local fields, $\hat{\vect{h}}$, are the solutions of the self-consistency equations

\begin{equation}
\frac{\partial \phi(\vect{J},\vect{h})}{\partial h_i} = m_i, \quad \frac{\partial \phi(\vect{J},\vect{h})}{\partial J_{ij}} = c_{ij}, \qquad i,j=1,\ldots,n.
\label{SK_mle}
\end{equation}

\noindent The universal codes for the SK model then reads as

\begin{equation}
\bar{P}(\hat{{s}}) = \exp \left[ N \left( \sum_i \hat{h}_i m_i + \sum_{i<j} \hat{J}_{ij} c_{ij} - \phi(\hat{\vect{J}}, \hat{\vect{h}}) \right) -\bar{\mathcal{R}} \right].
\label{SK_UC}
\end{equation}

\noindent However, unlike for the Dirichlet model and the paramagnet model, the UC partition function, $e^{\bar{\mathcal{R}}}$, for the SK model is analytically intractable (For SK models which possess some particular structures, a calculation of the UC partition function has been done in~ \cite{beretta2017stochastic}).
To this, we resort to a Markov chain Monte Carlo (MCMC) approach to sample the universal codes (See Appendix \ref{MCMC}).
Figure \ref{fig:uc_skandrbm}A and C shows the properties of the typical samples drawn from the universal codes of the SK model in Equation \eqref{SK_UC}.

\begin{figure}[!ht]
\centering
\includegraphics[width=1.0\textwidth]{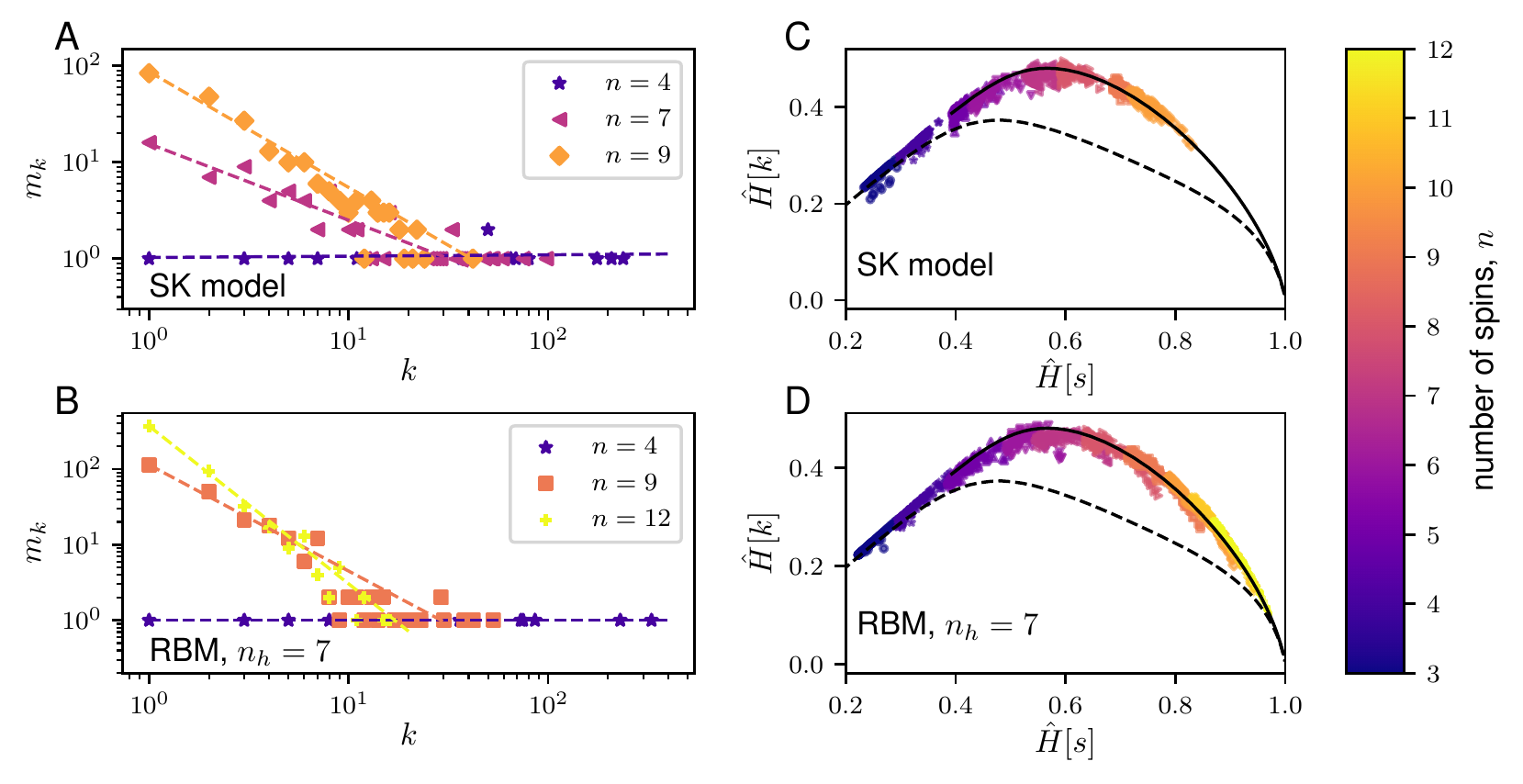}
\caption{\label{fig:uc_skandrbm}{Properties of typical samples for the NML codes of two graphical models: the Sherrington-Kirkpatrick (SK) model and the restricted Boltzmann machine (RBM)}.
Left panels ({\bf A},{\bf C}) show plots of the degeneracy, $m_k$, of the frequency, $k$, for representative typical samples generated from the NML codes for the SK model ({\bf A}) and the RBM given a number of hidden variables, $n_h=7$ ({\bf B}) for different number of (visible) spins, $n$.
The corresponding dashed lines show the best-fit lines.
On the other hand, right panels ({\bf B},{\bf D}) show plots of the $\hat{H}[k]$ versus $\hat{H}[s]$ of the typical samples drawn from the NML codes for the SK model ({\bf B}) and the RBM with $n_h=7$ ({\bf D}) for $N=10^3$ and for varying number of spins, $n$ ranging from $3$ to $12$.
Given $N$ and $n$ of a graphical model, we compute the $\hat{H}[k]$ and $\hat{H}[s]$ for 100 samples drawn from the respective NML codes through a Markov chain Monte Carlo (MCMC) approach (see Appendix \ref{MCMC}).
Note that for the RBM, varying $n_h$ do not qualitatively affect the observations made in this paper.
As before, the $\hat{H}[k]$ and $\hat{H}[s]$ are normalized by $\log N$ and the typical NML samples are compared against maximally informative samples (solid black line) and random samples (dashed black line) as described in Figure \ref{fig:uc_dirichlet}.}
\end{figure}

\subsubsection{Restricted Boltzmann Machines}
We consider a restricted Boltzmann machine (RBM) wherein one has a layer composed of $n_v$ independent visible boolean units, $\vect{v} = (v_1, \ldots, v_{n_v})$, which are interacting with $n_h$ independent hidden boolean units, $\vect{h}  = (h_1, \ldots, h_{n_h})$, in another layer where $v_i, h_i = 0,1$.
The probability distribution can be written down as

\begin{equation}
P(\vect{v}, \vect{h} | \vect{\theta} = (\vect{a}, \vect{b}, \vect{w}) ) = \frac{1}{Z(\vect{\theta})} \exp \left(\sum_{i=1}^{n_v} a_i v_i + \sum_{j=1}^{n_h} b_j h_j + \sum_{i=1}^{n_v} \sum_{j=1}^{n_h} v_i w_{ij} h_j \right)
\end{equation}

\noindent where the partition function

\begin{equation}
Z(\vect{\theta}) = \sum_{v_1 = 0,1} \cdots \sum_{v_{n_v} = 0,1} \sum_{h_1 = 0,1} \cdots \sum_{h_{n_h} = 0,1} \exp \left(\sum_{i=1}^{n_v} a_i v_i + \sum_{j=1}^{n_h} \left( b_j + \sum_{i=1}^{n_v} v_i w_{ij} \right) h_j \right)
\end{equation}

\noindent is a function of the parameters, $\vect{\theta}$, with $w_{ij}$ is the interaction strength between $v_i$ and $h_j$, $\vect{a}$ and $\vect{b}$ are the local fields acting on the visible $\vect{v}$ and hidden $\vect{h}$ units respectively.
Because the hidden units, $\vect{h}$, are mutually independent, we can factorize and then marginalize the sum over the hidden variables, $\vect{h}$, to obtain the distribution of a single observation, $\vect{v}$, as

\begin{equation}
P( \vect{v} | \vect{\theta} ) = \frac{ 1 }{ Z ( \vect{\theta} ) } \exp \left[ \sum_{i=1}^{n_v} a_i v_i + \sum_{j=1}^{n_h} \log 2 \cosh \left( \sum_{i=1}^{n_v} v_i w_{ij} + b_j  \right) \right].
\end{equation}

\noindent Then, the probability distribution for a sample, $\hat{\vect{v}} = (\vect{v}^{(1)}, \ldots, \vect{v}^{(N)})$, of $N$ observations is simply

\begin{equation}
P( \hat{\vect{v}} | \vect{\theta} ) = \prod_{k=1}^N p( \vect{v}^{(k)} | \vect{\theta} ).
\end{equation}

\noindent The parameters, $\hat{\vect{\theta}}$, can be estimated by maximizing the likelihood using the Contrastive Divergence (CD) algorithm~\cite{hinton2002training,hinton2006reducing} (see Appendix \ref{RBM_CD}).
Once the maximum likelihood parameters, $\hat{\vect{\theta}}$, have been inferred, then the universal codes for the RBM can be built as

\begin{equation}
\bar{P} ( \hat{\vect{v}} ) = e^{-\bar{\mathcal{R}}} \prod_{k=1}^N p( \vect{v}^{(k)} | \hat{\vect{\theta}} ).
\label{RBM_UC}
\end{equation}

In addition, like in the SK model, the UC partition function, $e^{\bar{\mathcal{R}}}$, for the RBM cannot be solved analytically.
To this, we also resort to a MCMC approach to sample the universal codes (See Appendix~ \ref{MCMC}). Figure \ref{fig:uc_skandrbm}B and D shows the properties of the typical samples drawn from the universal codes of the RBM in Equation \eqref{RBM_UC}.

Taken together, we see that even for models that incorporate interactions, the typical samples of the NML {\em i)} have broad frequency distributions and {\em ii)} they achieve values of $\hat{H}[k]$ close to the maximum, given $\hat{H}[s]$.
Due to computational constraints, we only present the results for $N=10^3$ however, we expect that increasing $N$ will only shift the NML towards the well-sampled regime.

\subsection{Large Deviations of the Universal Codes Exhibit Phase Transitions}\label{sec:atypical}

In this section, we focus on the distribution of the resolution $\hat{H}[s]$ for samples $\hat s$ drawn from $\bar P$. We note that

\[
\hat H[{s}] = \frac{1}{N} \sum_{i=1}^N \log \frac{k_{s^{(i)}}}{N}
\]

\noindent has the form of an empirical average. Hence, we expect it to attain a given value for typical samples drawn from $\bar P$. This also suggests that the probability  to draw samples with resolution $\hat{H}[s]=E$ different from the typical value has the large deviation form $P\{\hat{H}[s]=E\}\sim e^{-NI(E)}$, to leading order for $N\gg 1$. In order to establish this result and to compute the function $I(E)$, as in \cite{mezard2009information} and \cite{filiasi2014concentration}, we observe that

\begin{align}
{P \{ \hat{H}[s] = E \} }&{= \sum_{\hat{s}} \bar{P}(\hat{s}) \delta\left( \hat{H}[s] - E \right) } \\
											&{= \int_{-\infty}^{\infty} \frac{Ndq}{2 \pi} \sum_{\hat{s}} \bar{P}(\hat{s}) e^{iq N (\hat{H}[s] - E)},
\label{app3:eq1}
}
\end{align}

\noindent where we used the integral representation of the $\delta$ function and $\bar{P}(\hat{s})$ is the NML distribution in Equation \eqref{NML_prob}. Upon defining

\begin{equation}
{\sum_{\hat{s}} \bar{P}(\hat{s}) e^{iq N \hat{H}[s]} = e^{N \phi(iq)},
\label{app3:assumption}
}
\end{equation}

\noindent let us assume, as in the G\"artner--Ellis theorem \cite{mezard2009information}, that $\phi(iq)$ is finite for $N \gg 1$ for all $q$ in the complex plane. Then Equation (\ref{app3:eq1}) can be evaluated by a saddle point integration

\begin{align}
{P \{ \hat{H}[s] = E \} }&{= \int_{-\infty}^{\infty} \frac{N d\alpha}{2 \pi} e^{-N[i\alpha E-\phi(i\alpha)]}}\\
											&{\sim e^{-N[\beta E-\phi(\beta)]},
\label{app3:integral2}
}
\end{align}

where we account only for the leading order. $\beta$ is related to the saddle point value $q^* = -i \beta$ that dominates the integral and it is given by the solution of the saddle point condition

\begin{equation}
{E = \frac{d}{d\beta} \phi(\beta).
\label{app3:saddle}
}
\end{equation}

\begin{figure}[ht]
\centering
\includegraphics[width=0.8\textwidth]{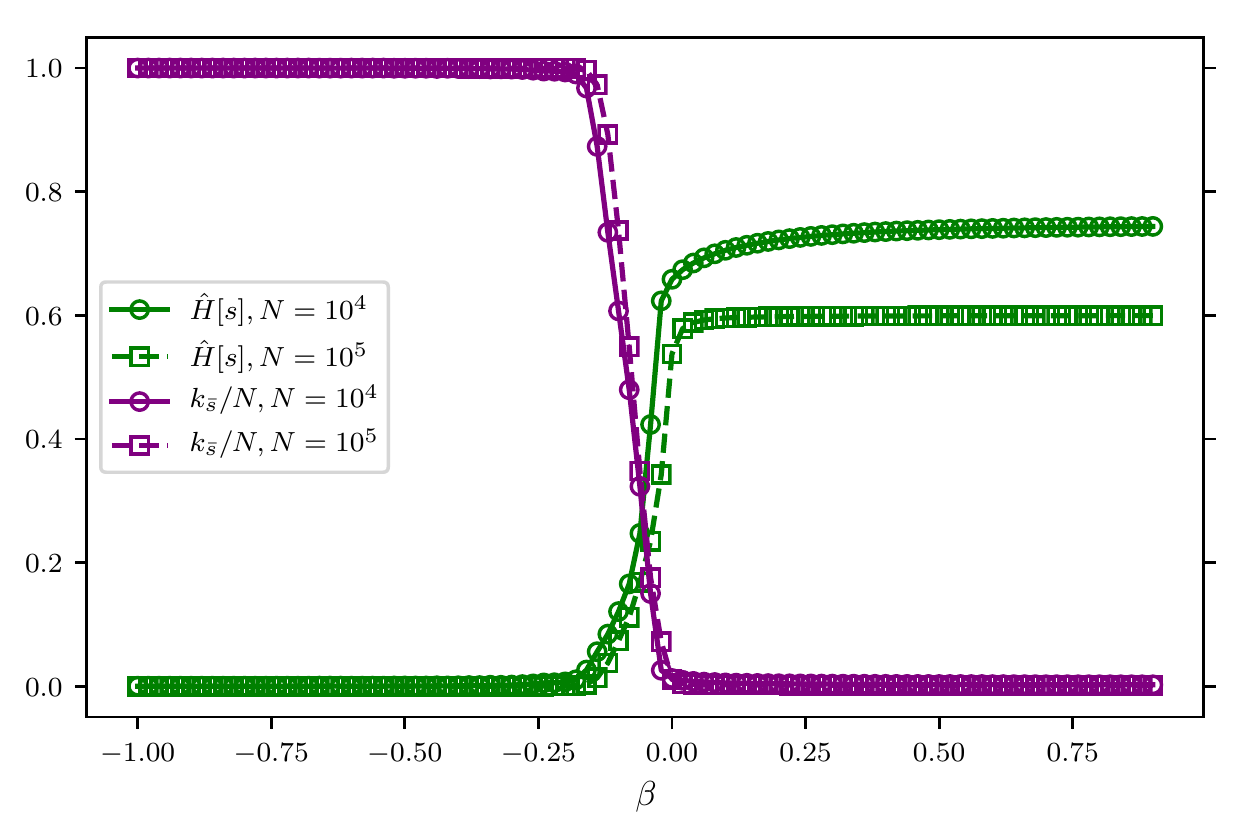}
\caption{\label{fig:ldtDirichlet} {Typical realizations of large deviations from the NML code of the Dirichlet model}. 
For a fixed parameter, $\beta$ ranging from $\beta=-1$ to $\beta=1$, samples are obtained from $\bar{P}_\beta$ in Equation \eqref{ldt_probability} for varying length of the dataset, $N$ ($N=10^4$ in solid lines with circle markers and $N=10^5$ in dashed lines with square markers).
The resolution $\hat{H}[s]$ normalized by $\log N$ (in green lines) and the maximal frequency $k_{\bar{s}}$ normalized by $N$ (in purple lines) are calculated as an average over 100 realizations of $\bar{P}_\beta$ given $\beta$.
The point $\beta=0$ corresponds to the typical samples that are realized from the Dirichlet NML code in Equation \eqref{pofk_dirichlet}.}
\end{figure}

\noindent Equation (\ref{app3:integral2}) shows that the function $I(E)$ is the Legendre transform of $\phi(\beta)$, i.e., 

\begin{equation}
{I(E) = - \beta E + \phi(\beta)}
\end{equation}

\noindent with $\beta(E)$ given by the condition (\ref{app3:saddle}), as in the G\"artner--Ellis theorem \cite{mezard2009information}. Further insight and a direct calculation from the definition in Equation \eqref{app3:assumption} reveals that Equation (\ref{app3:saddle}) can also be written as

\begin{equation}
{  E =\sum_{\hat s}\bar P_\beta(\hat s)\hat H[s]
\label{app3:saddle2}
}
\end{equation}

\noindent which is the average of $\hat H[s]$ over a ``tilted'' probability distribution \cite{mezard2009information}

\begin{equation}
{\bar P_\beta(\hat{s})=\bar P(\hat{s})e^{N\left\{\beta \hat{H}[s]-\phi(\beta)\right\}},
\label{ldt_probability}
}
\end{equation}

\noindent hence $\beta$ arises as the Lagrange multiplier enforcing the condition $\hat H[s]=E$. 
Conversely, when $\beta(E)$ is fixed by the condition Equation (\r
ef{app3:saddle2}), samples drawn from $\bar P_\beta$ have $\hat H[s]\simeq E$. In other words, $\bar P_\beta$ describes how large deviations with $\hat H[s]= E$ are realized.
Therefore, typical samples that realize such large deviations can be obtained by sampling the distribution $\bar P_\beta(\hat{s})$ in Equation \eqref{ldt_probability}.
Figure \ref{fig:ldtDirichlet} show that, for Dirichlet models, samples obtained from $\bar P_\beta$ exhibit a sharp transition at $\beta=0$.
The resolution (see green lines in Figure \ref{fig:ldtDirichlet}) sharply vanishes for negative values of $\beta$ as a consequence of the fact that the distribution {\em localizes} to samples where almost all outcomes coincide, i.e., $s_i=\bar s$.
This is evidenced by the fact that the maximal frequency $k_{\bar s}=\max_s k_s$ approaches $N$ very fast (see purple lines in Figure \ref{fig:ldtDirichlet}).
In other words, $\beta=0$ marks a {\em localization} transition where the symmetry between the states in $\chi$ is broken, because one state $\bar s$ is sampled an extensive number of times $k_{\bar s}\propto N$.

One direct way to see this is to consider the Dirichlet model and use the ``tilted'' distribution in Equation \eqref{ldt_probability} to compute the distribution 

\[
q_\beta(k | z) = \frac{1}{\mathcal{N}(z)} \frac{k^{(1-\beta)k}e^{- (1+z)k} }{k!}.
\]

\noindent of $k_s$ following the same steps leading to Equation (\ref{qofk_dirichlet}), where again $z$ is fixed by the condition
$\sum_k q_\beta(k | z)k=N/S$.
For $\beta\ge 0$, we again find, as in Equation (\ref{eq:pq}), that $k_s$ can be considered as independent draws from the same distribution $q_\beta(k | z)$.
For $\beta<0$, we find that the distribution $q_\beta(k | z)$ develops a sharp maximum at $k=N$ indicating that, as mentioned above, the sample concentrates on one state $\bar s$.

This behavior is generic whenever the underlying model $f(s|\theta)$ itself localizes for certain values $\bar \theta$ of the parameters, i.e., when $f(s|\bar\theta)=\delta_{s,\bar s}$.
In order to see this, notice that, in general, we can write

\begin{equation}
f ( \hat{s} | \hat{\theta}(\hat{s}) ) = \prod_s f (s | \hat{\theta}(\hat{s}) )^{k_s}.
\label{factorization}
\end{equation}

\noindent Thus, by inserting the identity $e^{-N\hat{H}[s] + N\hat{H}[s]}$, the NML distribution in Equation \eqref{NML_prob} can be re-cast as

\begin{equation}
\bar P(\hat{s})=e^{- N\hat{H}[s] - ND_{KL}(\hat p||\hat\theta)-\bar{\mathcal{R}}}
\label{recast_NML}
\end{equation}

\noindent where $\hat p_s=k_s/N$ is the empirical distribution and 

\begin{equation}
D_{KL}(\hat{p}||\hat{\theta})=\sum_s \hat{p}_s \log \frac{ \hat{p}_s }{ f( s | \hat{\theta}(\hat{s}) ) }
\end{equation}

\noindent is a Kullback-Leibler divergence.

Now, we observe that

\begin{align}
e^{N\phi(\beta)} & = e^{-\bar{\mathcal{R}}}\sum_{\hat{s}}e^{-(1-\beta)N\hat{H}[s]-ND_{KL}(\hat p||\hat\theta)} \label{eqC} \\
 & \ge e^{-\bar{\mathcal{R}}}\sum_{\hat{s}}e^{-(1-\beta)N\hat{H}[s]-ND_{KL}(\hat p||\theta_0)} \label{eqB} \\
 & = e^{-\bar{\mathcal{R}}}\label{eqA}
\end{align}

\noindent where the inequality in Equation \eqref{eqB}  derives from the fact that $\hat\theta(\hat{s})$, the maximum likelihood estimator for sample $\hat{s}$, is replaced by a generic value $\theta_0$and consequently, $D_{KL}(\hat p||\hat\theta) \le D_{KL}(\hat p||\theta_0)$.
The equality in Equation \eqref{eqA} , instead, derives from the choice $\theta_0=\bar \theta$ such that $f(s|\bar\theta)=\delta_{s,\bar s}$.
Under this choice, only the term corresponding to ``localized'' samples where $s^{(i)}=s_0$ for all points in the sample, survive in the sum on $\hat{s}$.
For such localized samples, $\hat{H}[s]=D_{KL}(\hat p||\theta_0)=0$, hence Equation \eqref{eqA} follows.

Because of the logarithmic dependence of the regret $\bar{\mathcal{R}}$ on $N$ (see Equation \eqref{parametric_complexity}), Equation \eqref{eqA} implies that, for all $\beta$,

\begin{equation}
\phi(\beta) \ge \bar{\mathcal{R}}/N \simeq 0
\end{equation}

\noindent for $N\gg 1$. Given that $\hat{H}[s] \ge 0$ in Equation \eqref{app3:saddle2}, then $E \ge 0$ and therefore, Equation \eqref{app3:saddle} implies that $\phi(\beta)$ is a non-decreasing function of $\beta$. In addition,  $\phi(0)=0$ by Equation \eqref{app3:assumption}.
Taken together, these facts require that $\phi(\beta)=0$ for all values $\beta \le 0$.
On the other hand, for $\beta>0$, the function $\phi(\beta)$ is analytic with all finite derivatives, which corresponds to higher moments of $\hat H[s]$ under $\bar P_\beta$. 
Therefore, $\beta=0$, which corresponds to the typical behavior of the NML, coincides with a second order phase transition point because the function $\phi(\beta)$ exhibits a discontinuity in the second derivative.
In terms of $\bar P_\beta(\hat{s})$, the phase transition separates a region ($\beta \ge 0$) where all samples $\hat{s}$ have a finite probability from a region ($\beta < 0$) where only one sample, the one with $s^{(i)}=\bar s, \forall i$, has non-zero probability and $\hat{H}[s]=0$.

The phase transition is a natural consequence of the fact that NML provide efficient coding of samples generated from $f(s|\theta)$.
It states that codes $\bar P_\beta$ that achieve a compression different from the one achieved by the NML only exist for higher coding costs.
Codes with lower coding cost only describe non-random samples that correspond to deterministic models $f(s|\bar\theta)=\delta_{s,\bar s}$.

\section{Discussion}

The aim of this paper is to elucidate the properties of efficient representations of data corresponding to universal codes that arise in MDL. 
Taking NML as a generative model, we find that typical samples are characterized by broad frequency distributions and that they achieve values of the relevance which are close to the maximal possible $\hat H[k]$.

In addition, we find that samples generated from NML are critical in a very precise sense.
If we force NML to use less bits to encode samples, then the code localizes on deterministic samples.
This is a consequence of the fact that if there were codes that required fewer bits, then NML would not be optimal. 

This contributes to the discussion on the ubiquitous finding of {\em statistical criticality} \cite{munoz2018colloquium,mora2011biological} by providing a clear understanding of its origin.
It suggests that {\em statistical criticality} can be related to a precise second order phase transition in terms of large deviations of the coding cost.
This phase transition separates random samples that span a large range of possible outcomes (the set $\chi$ in the models discussed above) from deterministic ones, where one outcome occurs most of the time.
The phase transition is accompanied by a {\em spontaneous symmetry breaking} in the permutation between samples.
The frequencies of outcomes in the symmetric phase ($\beta\ge 0$) are generated as independent draws from the same distribution, that is sharply peaked for $\beta>0$ as can be checked in the case of the Dirichlet model.
Instead, for $\beta<0$, only one state is sampled.
In the typical case, $\beta=0$, the symmetry between outcomes is weakly broken, as there are outcomes that occur more frequently than others.
At $\beta = 0$, the samples maintain the maximal discriminative power over outcomes.
This type of phase transitions in large deviations is very generic, and it occurs in large deviations whenever the underlying distribution develops fat tails (see e.g., \cite{filiasi2014concentration}). 

This leads to the conjecture that broad distributions arise as a consequence of efficient coding. 
More precisely, broad distributions arise when the variables sampled are relevant, i.e., when they provide an optimal representation. 
This is precisely the point which has been made in \cite{marsili2013sampling,haimovici2015criticality,cubero2018minimally}.
The results in the present paper add a new perspective whereby maximally informative samples can be seen as universal codes.


\appendix
\section{Derivation for the Parametric Complexity}
\label{saddle}
In order to compute the parametric complexity, given in Equation \eqref{UC_regret}, let us consider the integral $\int d\theta f( \hat{s} | \theta ) g(\theta)$ for a generic function $g(\theta)$. For $N\gg 1$, the integral is dominated by the point $\theta=\hat\theta(\hat s)$ that maximizes $\log f(\hat s|\theta)$, and it can be computed by the saddle point method.
Performing a Taylor expansion around the maximum likelihood parameters, $\hat{\theta}(\hat{s})$, one finds (up to leading orders in $N$)

\begin{align}
\log f( \hat{s} | \theta ) &= \log f( \hat{s} | \hat{\theta}(\hat{s}) ) - \frac{1}{2} \sum_{i,j} N (\theta_i-\hat\theta_i) I_{i,j}(\hat\theta) (\theta_j-\hat\theta_j) + \mathcal{O}((\theta-\hat\theta)^3).
\end{align}

\noindent where

\begin{align}
I_{i,j}(\hat{\theta}) &= - \frac{1}{N} \frac{\partial^2 \log f ( \hat{s} | \theta)}{\partial \theta_i \partial \theta_j}\\
& = -\sum_{s\in\chi} \frac{k_s}{N} \frac{\partial^2 \log f (s | \theta)}{\partial \theta_i \partial \theta_j}.\label{appA:eqn1}
\end{align}

\noindent Note that for exponential families, the Hessian of the log-likelihood is independent of the data, and hence it coincides with the Fisher Information matrix \cite{balasubramanian2005mdl}

\begin{equation}
I_{i,j}(\theta) = -\sum_{s\in\chi}f(s|\theta)\frac{\partial^2 \log f (s | \theta)}{\partial \theta_i \partial \theta_j}\label{appA:eqn2}
\end{equation}

\noindent The integral can then be computed by Gaussian integration, as

\begin{align}
\int d\theta f( \hat{s} | \theta ) g(\theta) & \simeq f(\hat{s} | \hat{\theta}(\hat{s})) g(\hat{\theta}(\hat{s})) \int d\theta e^{-\frac{N}{2} \sum_{i,j} (\theta_i -\hat\theta_i) I_{ij}(\hat{\theta})(\theta_j-\hat\theta_j)}\label{integral_equationB}\\
& = f(\hat{s} | \hat{\theta}(\hat{s})) g(\hat{\theta}(\hat{s})) \left( \frac{2 \pi}{N}\right)^{\frac{k}{2}} \frac{1}{\sqrt{\det {I}(\hat{\theta)}}}.
\label{integral_equationA}
\end{align}

\noindent where $k$ is the number of parameters.
If we choose $g(\theta)$ to be

\begin{equation}
g(\theta) = \left( \frac{N}{2 \pi} \right)^{\frac{k}{2}} \sqrt{ \det I(\theta) }
\end{equation}

\noindent and take a sum over all samples $\hat s$ on both sides of Equation \eqref{integral_equationB}, Equation (\ref{integral_equationA}) becomes

\begin{align}
\sum_{\hat{s}} f(\hat{s} | \hat{\theta}(\hat{s})) &\simeq \sum_{\hat{s}} \left( \frac{N}{2 \pi} \right)^{\frac{k}{2}} \int d\theta f( \hat{s} | \theta ) \sqrt{ \det I(\theta) } \\
&= \left( \frac{N}{2 \pi} \right)^{\frac{k}{2}} \int \sqrt{ \det I(\theta) } d\theta .
\end{align}

\noindent Hence, the parametric complexity, $\bar{\mathcal{R}} = \log \sum_{\hat{s}} f(\hat{s} | \hat{\theta}(\hat{s}))$, is asymptotically given by Equation \eqref{parametric_complexity} when $N\gg 1$.

Notice also that $\bar{P}(\hat{s})$ induces a distribution over the space of parameters $\theta$. With the choice

\begin{equation}
g(\theta) = \left( \frac{N}{2 \pi} \right)^{\frac{k}{2}} \sqrt{ \det I(\theta) }\delta(\theta-\theta_0),
\end{equation}

\noindent the same procedure as above shows that

\begin{align}
\sum_{\hat{s}} \bar{P}(\hat{s}) \delta\left(\hat\theta(\hat s)-\theta_0\right) &=e^{-\bar{\mathcal{R}}} \sum_{\hat{s}} f(\hat{s} | \hat{\theta}(\hat{s})) \delta\left(\hat\theta(\hat s)-\theta_0\right) \\
&=e^{-\bar{\mathcal{R}}} \left( \frac{N}{2 \pi} \right)^{\frac{k}{2}} \sqrt{ \det I(\theta_0) } \\
&=\frac{\sqrt{ \det I(\theta_0) }}{\int d\theta \sqrt{ \det I(\theta) }}
\end{align}

\noindent which is the Jeffreys prior.

\section{Calculating the Parametric Complexity}
In this section, we calculate the parametric complexity for the Dirichlet model for $\rho = N/S \gg 1$ where $N$ is the number of observations in the sample $\hat{s}$ and $S$ is the size of the state space $\chi$ and the paramagnetic Ising model.

\subsection{Dirichlet Model}\label{dirichlet_partition}
In the regime where $\rho \gg 1$ and $k$ large such that we can employ Stirling's approximation, $k! = \sqrt{2 \pi k} k^k e^{-k}$, the normalization can be calculated as

\begin{align}
\sum_{k=0}^{\infty} \frac{k^{k}e^{- k} e^{- z^*(\rho) k}}{k!} &\approx \sum_{k=0}^{\infty} \frac{e^{- z^*(\rho) k}}{\sqrt{2 \pi k}} \\
&= \int_0^\infty \frac{e^{- z^*(\rho) k} dk}{\sqrt{2 \pi k}} \\
&= \frac{1}{\sqrt{2 \pi}} \sqrt{\frac{\pi}{z^*(\rho)}} \\
&= \frac{1}{\sqrt{2 z^*(\rho)}}.
\end{align}

\noindent Similarly, we can also calculate

\begin{align}
\sum_{k=0}^{\infty} \frac{k^{k+1}e^{- k} e^{- z^*(\rho) k}}{k!} &\approx \sum_{k=0}^{\infty} \frac{k e^{- z^*(\rho) k}}{\sqrt{2 \pi k}}  \\
&= \int_0^\infty \sqrt{\frac{k}{2 \pi}} e^{- z^*(\rho) k} dk \\
&= \frac{1}{\sqrt{2 \pi}} \frac{\sqrt{\pi}}{2(z^*(\rho))^{\frac{3}{2}}} \\
&= \frac{1}{(2 z^*(\rho))^{\frac{3}{2}}}
\end{align}

\noindent and thus, the saddle point value $z^*$ can now be evaluated as

\begin{equation}
z^*(\rho) \simeq \frac{1}{2 \rho}.
\end{equation}

In the same regime, given the determinant $\det I(\theta)$ of the Fisher information matrix for the Dirichlet model,

\begin{equation}
\det I(\theta) = \prod_{s \in \chi} \frac{1}{\theta_s}
\end{equation}

\noindent the parametric complexity can be approximated as

\begin{align}
e^{\bar{\mathcal{R}}} &\simeq \left( \frac{N}{2 \pi} \right)^{ (S-1)/2} \int d\theta \sqrt{\det I(\theta)} \\
   &= \left( \frac{N}{2 \pi} \right)^{(S-1)/2} \frac{\Gamma(\frac{1}{2})^S}{\Gamma(\frac{S}{2})} \\
   &\simeq  \frac{e^{ \frac{S}{2} ( 1 + \log \rho ) }}{\sqrt{2 \rho}}
 \end{align}

\noindent which, together with Equation \eqref{dirichlet_regretA} and the fact and the variance $\langle k^2 \rangle_{z^*} - \langle k \rangle_{z^*}^2 = 2 \rho^2$,
implies that $\Phi(z^*(\rho)) = \frac{1}{2} ( 1 + \log \rho )$.

\subsection{Paramagnet Model}\label{paramagnet_partition}
The parametric complexity for the paramagnetic Ising model, given $\bar{P}(m)$ in Equation \eqref{UC_paramagnet_pofm}, is given by

\begin{equation}
e^{\bar{\mathcal{R}}} = \sum_{M=-N}^{N} \binom{N}{\frac{N-M}{2}} e^{ \left[ M \tanh^{-1} (M/N) - N \log \left( 2 \cosh \left( \tanh^{-1} (M/N)  \right) \right) \right] }.
\end{equation}

\noindent where $M=-N,-N+2,\ldots,N-2,N$ runs on $N+1$ values. When  $N \gg 1$, the magnetization, $m=M/N$, can be treated as a continuous variable and consequently, the sum can be approximated as an integral: $\sum_M\ldots \simeq \frac N 2 \int_{-1}^1dm\ldots$. Hence, by using the identities $\tanh^{-1}(m) = \frac{1}{2} \log \frac{1+m}{1-m}$, $\cosh \left( \tanh^{-1}(m) \right) = \frac{1}{\sqrt{1-m^2}}$ and $K!\simeq K^Ke{-K}\sqrt{2\pi K}$, one finds that

\begin{align}
e^{\bar{\mathcal{R}}} &\simeq \frac{N}{2}\int_{-1}^{1} \frac{1}{\sqrt{2\pi N(1-m^2)}}\\
	&=\sqrt{ \frac{\pi N}{ 2} }.
\end{align}

\section{Simulation Details}
\subsection{Sampling Universal Codes Through Markov Chain Monte Carlo}\label{MCMC}
Unlike the Dirichlet model and the independent spin model, analytic calculations for the Sherrington-Kirkpatrick (SK) model and the restricted Boltzmann machine (RBM) are generally not possible, because the partition function $Z$, and consequently, the UC partition function $e^{\mathcal{\bar{R}}}$, is computationally intractable. In order to sample the NML for these graphical models, we turn to a Markov chain Monte Carlo (MCMC) approach in which the transition probability, $P(\hat{s} \to \hat{s}')$, can be built using the following heuristics:
\begin{enumerate}
\item Starting from the sample, $\hat{s}$, we calculate the maximum likelihood estimates, $\hat{\vect{\theta}}(\hat{s})$, of the parameters of the model, $p(\hat{s} | \theta)$ by either solving Equation \eqref{SK_mle} for the SK model or by Contrastive Divergence (CD$_{\kappa}$)~\cite{hinton2002training,hinton2006reducing} for the RBM (see Appendix \ref{RBM_CD}). \label{step1}
\item We generate a new sample, $\hat{s}'$ from $\hat{s}$ by flipping a spin in randomly selected $r$ points $s^{(i)}$ of the sample. The number of selected spins, $r$, must be chosen carefully such that $r$ must be large enough to ensure faster mixing but small enough so the new inferred model, $p(\hat{s}' | \theta)$, is not too far from the starting model, $p(\hat{s} | \theta)$. \label{step2}
\item The maximum likelihood estimators, $\hat{\vect{\theta}}(\hat{s}')$ for the new sample are calculated as in Step \ref{step1}. \label{step3}
\item Compute

\begin{equation}
\Delta E = \log p(\hat{s}' | \hat{\vect{\theta}}(\hat{s}')) - \log p(\hat{s} | \hat{\vect{\theta}}(\hat{s}))
\end{equation}

\noindent and accept the move $\hat{s} \to \hat{s}'$ with probability $\min\left( e^{N \Delta E}, 1 \right)$. \label{step4}
\end{enumerate}

\subsection{Estimating RBM Parameters Through Contrastive Divergence}\label{RBM_CD}
Given a sample, $\hat{\vect{v}} = (\vect{v}^{(1)}, \ldots, \vect{v}^{(N)})$, of $N$ observations, the log-likelihood for the restricted Boltzmann machine (RBM) is given by

\begin{equation}
\log \mathcal{L}(\vect{\theta}) = \sum_{k = 1}^N \log \sum_{\vect{h}} P(\vect{v}^{(k)}, \vect{h} | \vect{\theta}).
\end{equation}

\noindent The inference of the parameters, $\vect{\theta}$, proceeds by updating $\vect{\theta}$ such that the log-likelihood, $\log \mathcal{L}(\theta)$, is maximized. This updating formulation for the parameters is given by

\begin{equation}
\Delta \theta = \frac{\epsilon}{N} \frac{\partial \log \mathcal{L}(\vect{\theta})}{\partial \theta}
\end{equation}

\noindent where $\epsilon$ is the learning rate parameter. The corresponding gradients for the parameters, $\vect{w}$, $\vect{a}$ and $\vect{b}$ can then be written down respectively as

\begin{align}
\frac{\partial \log \mathcal{L}(\vect{\theta})}{\partial w_{ij}} &= \sum_{k=1}^N \left[ \sum_{\vect{h}} v_i^{(k)} h_j P(\vect{v}^{(k)}, \vect{h} | \vect{\theta}) - \sum_{\vect{v}} \sum_{\vect{h}} v_i h_j P(\vect{v}, \vect{h} | \vect{\theta}) \right] \label{rbm_wij} \\
\frac{\partial \log \mathcal{L}(\vect{\theta})}{\partial a_i} &= \sum_{k=1}^N \left[ \sum_{\vect{h}} v_i^{(k)} P(\vect{v}^{(k)}, \vect{h} | \vect{\theta}) - \sum_{\vect{v}} \sum_{\vect{h}} v_i P(\vect{v}, \vect{h} | \vect{\theta}) \right] \\
\frac{\partial \log \mathcal{L}(\vect{\theta})}{\partial b_j} &= \sum_{k=1}^N \left[ \sum_{\vect{h}} h_j P(\vect{v}^{(k)}, \vect{h} | \vect{\theta}) - \sum_{\vect{v}} \sum_{\vect{h}} h_j P(\vect{v}, \vect{h} | \vect{\theta}) \right]
\end{align}

\noindent where the first terms denote averages over the data distribution while the second terms denote averages over the model distribution.

Here, we use the contrastive divergence (CD) approach which is a variation of the steepest gradient descent of $\mathcal{L}(\vect{\theta})$. Instead of performing the integration over the model distribution, CD approximates the partition function by averaging over distribution obtained after taking $\kappa$ Gibbs sampling steps away from the data distribution.

To do this, we exploit the factorizability of the conditional distributions of the RBM. In particular, the conditional probability for the forward propagation (i.e., sampling the hidden variables given the visible variables) from $\vect{v}$ to $h_j$ reads as

\begin{equation}
P( h_j = 1 | \vect{v}, \vect{\theta} ) = \frac{1}{1 + \exp\left( -b_j - \sum_i v_i w_{ij} \right)}.
\end{equation}

\noindent Similarly, the conditional probability for the backward propagation (i.e., sampling the visible variables from the hidden variables) from $\vect{h}$ to $v_i$ reads as

\begin{equation}
P( v_i= 1 | \vect{h}, \vect{\theta} ) = \frac{1}{1 + \exp\left( -a_i - \sum_j h_j w_{ij} \right)}.
\end{equation}

\noindent The Gibbs sampling is done by propagating a sample, $\vect{v}^{(k)} = \vect{v}^{(k)}(0)$, forward and backward $\kappa$ times: $\vect{v}^{(k)}(0) \to \vect{h}^{(k)}(0) \to \vect{v}^{(k)}(1) \to \ldots \to \vect{h}^{(k)}(\kappa-1) \to \vect{v}^{(k)}(\kappa) \to \vect{h}^{(k)}(\kappa)$. And thus, the Gibbs sampling approximates the gradient in Equation \eqref{rbm_wij} as

\begin{equation}
\frac{\partial \log \mathcal{L}(\vect{\theta})}{\partial w_{ij}} = \sum_{k=1}^N \left[ v_i^{(k)}(0)h_j^{(k)}(0) - v_i^{(k)}(\kappa) h_j^{(k)}(\kappa) \right].
\end{equation}

In the CD approach, each parameter update for a batch is called an epoch. While larger $\kappa$ approximates well the partition function, it also induces an additional computational cost. To find the global minimum more efficiently, we randomly divided the samples into groups of mini-batches. This approach introduces stochasticity and consequently reduces the likelihood of the learning algorithm to be confined in a local minima. However, a mini-batch approach can result in data-biased sampling. To circumvent this issue, we adopted the Persistent CD (PCD) algorithm where the Gibbs sampling extends to several epochs, each using different mini-batches. In the PCD approach, the initial visible variable configuration, $\vect{v}^{(k)}(0)$, was set to random for the first mini-batch, but the final configurations, $(\vect{v}^{(k)}(\kappa), \vect{h}^{(k)}(\kappa))$, of the current batches become the initial configuration for the next mini-batches. In this paper, we performed Gibbs sampling at $\kappa=10$ steps where we update the parameters, $\vect{\theta}$, are updated at $2500$ epochs at a rate $\epsilon=0.01$ with $200$ mini-batches per epochs. For other details regarding inference of parameters of the RBM, we refer the reader to \cite{hinton2002training,hinton2006reducing}.

\subsection{Source Codes}\label{codes}
All the calculations in this manuscript were done using personalized scripts written in Python 3. The source codes  are accessible online (https://github.com/rcubero/UniversalCodes).

\end{document}